\documentclass[%12pt,         %
aps,                    %  American Physical Society
prd,                    %  Physical Review Letters
superscriptaddress,    %  Authors' addresses linked with superscripts
nofootinbib,            %  Does not treat footnotes as references
amsmath,               %
amssymb,               %
floatfix]               %  Fixes float errors
{revtex4}               %  REVTEX 4 Package used

\usepackage{float}
\usepackage{amssymb}
\usepackage{graphicx}
\usepackage{amsmath}
\usepackage{hyperref}
\usepackage{amsfonts}
\newcommand{\fr}{\frac}

\newcommand{\nonu}{\nonumber}

\newcommand{\be}{\begin{equation}}
\newcommand{\ee}{\end{equation}}
\newcommand{\beqa}{\begin{eqnarray}}
\newcommand{\eeqa}{\end{eqnarray}}

\begin{document}

\title{Gravitational lensing in Weyl gravity}

\author{O\u{g}uzhan Ka\c{s}\i k\c{c}\i}
\email{oguzhan.kasikci@msgsu.edu.tr}
\affiliation{Department of Physics, Mimar Sinan Fine Arts University,
Bomonti, \.{I}stanbul 34380, Turkey}

\author{Cemsinan Deliduman}
\email{cemsinan@msgsu.edu.tr}
\affiliation{Department of Physics, Mimar Sinan Fine Arts University,
Bomonti, \.{I}stanbul 34380, Turkey}
\affiliation{National Astronomical Observatory of Japan, 2-21-1 Osawa,
Mitaka, Tokyo 181-8588, Japan}

\begin{abstract}
We calculate the deflection angle of light from a distant source by a galaxy cluster in Weyl's conformal gravity. The general method of calculation is first applied to calculate the deflection angle in Schwarzschild-de Sitter (Kottler) spacetime. The deflection angle calculated in Kottler spacetime includes the contribution of the cosmological constant, which quantitatively agrees with one work and disagrees with many works in the literature. We then calculate the deflection angle in Mannheim-Kazanas spacetime in two conformally related coordinate systems and find that the result includes contributions from both the cosmological constant and the Mannheim-Kazanas parameter. There are conflicting results on the deflection angle for light in Weyl gravity in the literature. We point out a possible reason for the discrepancy between our work and the others. 
\end{abstract}

\maketitle

%%%%%%%%%%%%%%%%%%%%%%%%%%%%%%%%%%%%%%%%%%%%

\section{Introduction}

Bending of light by gravitational field was first predicted in Newtonian physics a long time ago \cite{Trimble2001,Valls1206}. However, the amount of bending due to the Sun, calculated within general relativity to be twice the Newtonian value, was the one that agreed with the observation in 1919 \cite{Eddington1919} and made Einstein the most famous physicist overnight. One important consequence of the gravitational bending of light is the gravitational lensing phenomenon \cite{Chwolson1924,Link1936,Einstein1936}, in which light from a distant object is bent by an intermediate massive object, that is, a gravitational lens, to create multiple images of the source. Gravitational lensing is a successful astronomical tool to obtain a great deal of information about the distance of the source, its brightness, and perhaps most importantly the mass distribution of the lensing object. If that object is a galaxy cluster, then gravitational lensing, together with x--ray observations, is an indispensable tool to measure the amount of mass in galactic constituents and the intergalactic gas. Those observations gave support to the dark matter paradigm that galaxy halos contain dark matter with much higher mass compared to the luminous baryonic mass making up the stars, as well as galactic and intergalactic gas. Existence of dark matter in galactic halos has also been the standard explanation of the phenomenon of flat galactic rotation curves.

Recent years have seen much activity testing the viability of alternative theories of gravity on astrophysical and cosmological phenomena without involving dark sectors. The dark matter paradigm, although very successful in explaining diverse astrophysical and cosmological phenomena, has compatibility issues with particle physics. Dark matter particles should belong to some theory beyond Standard Model; alas, there exists no model which is free of theoretical problems, and most importantly there are no observations reported by direct or indirect particle physics experiments (for the latest observational status, see Refs. \cite{Tanabashi2018,DMP}). The phenomenological success of Milgrom's MOND approach \cite{Milgrom1983,Famaey2012} together with the current (non) observational situation of dark matter particles make it imperative to search for an explanation of various astrophysical and cosmological phenomena in an alternative theory of gravity.

In a previous work \cite{DKY2015}, we determined the geometry in the outer region of galaxies in which stars move with almost the same rotational velocity irrespective of their distance from the galactic center. Constancy of the rotational velocity might be seen due to the existence of scale symmetry. Therefore, we found this geometry as a solution of Weyl gravity theory, which is the unique local scale symmetric metric theory of gravity. In Ref. \cite{DKY2015}, we also claimed that our solution for the outer region of galaxies should also hold for low-density regions up to the scale of galaxy clusters. To check the validity of that claim, in this paper, we analyze the gravitational lensing in Weyl gravity by calculating the deflection angle of light from a distant source by a galaxy cluster. 

There is some resistance in the scientific literature to accept the possibility of an alternative explanation of dark matter phenomena through the modification of the theory of gravitation. There exists an unfortunate widespread point of view that the dark matter paradigm was ``proven'' a long time ago \cite{Clowe2006} and alternative explanations are futile, even though there are numerous observational and theoretical works challenging the so called ``proof'' \cite{Mahdavi2007,Jee2012,Jee2014,McGaugh2016,Mastropietro2008,Lee2010}. Phenomenological successes of MOND and some other alternative theories of gravity have proven very difficult to carry on to the cosmological phenomena. Thus, even though Ref.\cite{DKY2015} has no error in its mathematical approach and soundness of its result, it was expected to explain much more than the mere phenomenon of flat rotation curves of galaxies. It was outside the scope of Ref. \cite{DKY2015} to resolve all the astrophysical and cosmological phenomena related to dark matter, which we intend to do in time as far as it is possible. It is unjust to expect a single paper to bring explanations to countless diverse observations. Thus, the present paper and also Ref. \cite{DKY2015} should be thought of as part of ongoing research to understand the relevance of Weyl gravity to natural phenomena, which has a history almost as long as one of general relativity \cite{Weyl1,Weyl2,Bach,Mann}.

This paper is organized as follows. In the next section, we summarize the previous works on the effect of the cosmological constant on gravitational lensing and the works on gravitational lensing in Weyl gravity. Then, in Sec. \ref{general}, we are going to describe the general formalism in two conformally related coordinate systems with two different methods of calculation. We will apply the general formalism first to Schwarzschild--de Sitter (Kottler) spacetime in Sec. \ref{SdS}, and then in Sec. \ref{MKS}, we will obtain the main result of this paper for the deflection angle of light in Weyl gravity.

%%%%%%%%%%%%%%%%%%%%%%%%%%%%%%%%%%%%%%%%%%%%

\section{Summary of previous works} \label{lit}

There are unfinished discussions in the literature on the contribution of the cosmological constant ($\Lambda$) and the Mannheim-Kazanas (MK) parameter ($\gamma$) of the MK solution \cite{MaKa,Mann} of Weyl gravity to the strong lensing formula. That $\Lambda$ contributes to the bending of light in the Kottler background was first proposed in Ref. \cite{Rindler2007}. Then, through a series of papers \cite{Ishak2008,Sereno2008,Schucker2009,Gibbons0808,Ishak2010,Rindler2010,Bhadra2010}, the contribution of $\Lambda$  to strong lensing is made more precise both conceptually and computationally. It is observed that the null geodesics being independent of $\Lambda$ \cite{Islam1983} does not imply that lensing phenomena are independent of $\Lambda$. Related to this discussion in a study involving dark matter \cite{Minic1601}, it is concluded that cold dark matter mass profiles contain information about $\Lambda$. 

There have also been some works \cite{Khriplovich2008,Park2008,Simpson2008,Arakida2012} questioning the validity of these results. In these papers, objections to the contribution of $\Lambda$ to the strong lensing formula can be grouped into three categories: 1) The Rindler-Ishak result \cite{Rindler2007} is for a static observer, and if the observer's motion or the cosmological Hubble flow is taken into account, the $\Lambda$-dependent terms would simply cancel in the final result \cite{Khriplovich2008,Park2008}. 2) As it was pointed out a long time ago in Ref. \cite{Islam1983} $\Lambda$ does not influence the null orbit equation, it can further be shown that $\Lambda$ can be absorbed into the definition of measurable quantities and hence does not influence the measurable quantities either \cite{Khriplovich2008,Park2008,Arakida2012}. 3) Any influence coming from  $\Lambda$ should be in higher-order terms, and lower-order influence observed in other works is due to the observer's motion \cite{Simpson2008}. These objections are well answered in Refs.\cite{Lake2013,Lake2016,Ishihara2016,He2017}, on which we are going to comment at the end of Sec. \ref{gamma0}.

There is also disagreement on the contribution of the MK parameter ($\gamma$) to the strong lensing formula. If one uses the MK solution of Weyl gravity to describe the galactic rotation curves, then the parameter $\gamma$ that multiplies the linear term in the gravitational potential turns out to be very small, but positive \cite{Mannheim1996,Mannheim2010}. Early works on strong lensing by galaxy clusters in the MK metric reported literally negative results that value of $\gamma$ should be negative for it to have positive contribution to the bending angle \cite{Walker1994,Edery9708,Nandi2010,Nandi2011}. Pireaux claimed \cite{Pireaux0408} that the MK choice of the conformal factor gives an incorrect value for $\gamma$. In a later work \cite{Sultana2010}, Sultana and Kazanas again found a formula which requires $\gamma$ to be negative, but their result suggested that the contribution of $\gamma$ is rather insignificant compared to the general relativistic contribution. Then, in the late works \cite{Sultana2013,Cattani2013,Cutajar1403,Potapov2016,Lim-Wang2017}, people used different ideas in the definition of the bending angle and found that $\gamma$ should be positive, getting rid of an apparent paradox. 

In this work, we are going to contribute to both of these discussions. First, it will be observed that our Weyl gravity solution \cite{DKY2015} is conformally equivalent to the MK solution, as any solution to Weyl gravity field equations should be \cite{MaKa}. So, we have a different conformal factor compared to the MK solution, which makes a difference only in the case of massive particle trajectories. Light trajectories do not distinguish conformally equivalent metrics; thus, our result for strong lensing is relevant for the discussion on the sign and the value of the MK parameter $\gamma$. Our result for the deflection angle calculated in the Kottler spacetime also includes a contribution from the cosmological constant $\Lambda$. This contribution, however, comes out rather differently than the works mentioned above. We believe that how $\Lambda$ and also $\gamma$ contribute depends strongly on how and at what point in the calculation of the deflection angle, the perturbative expansions in various quantities are performed. These quantities are mass $m$ of the gravitational lens, the cosmological constant $\Lambda$, and the MK parameter $\gamma$. In which order the perturbative expansions are made is very important. We find out that expansions first in $m$, then in $\gamma$, and finally in $\Lambda$ are the mathematically correct ones because otherwise one gets higher-order terms larger than the lower-order terms in perturbation expansions. This type of behavior is physically incorrect. Our result without the MK parameter for the deflection angle (in Kottler spacetime) agrees with the analysis done in Ref. \cite{Batic2015}, which has different result compared to Refs. \cite{Ishak2008,Sereno2008,Schucker2009,Gibbons0808,Ishak2010,Nandi2011,Rindler2010,Bhadra2010}. With the MK parameter, our result also differs from the results in the literature, partially agreeing only with the result presented in Ref. \cite{Lim-Wang2017}.

%%%%%%%%%%%%%%%%%%%%%%%%%%%

\section{General formalism} \label{general}

In this section, we describe the general formalism on a spherically symmetric static spacetime, first in the Schwarzschild-like polar-areal coordinates \cite{Parry2012} and then in conformally equivalent coordinates that we call ``Weyl gravity vacuum coordinates.'' For the null geodesics, the conformal transformation of the metric would not have any effect on the geodesic nor, thus, the deflection angle. By doing the calculation in two conformally related coordinate systems with two different methods and obtaining the same result, we seek to have some confidence in our result and the method used \cite{Will2003}. We are aware that to claim full coordinate independence a fully covariant calculation is needed.

\subsection{Polar-areal coordinates}

A general Schwarzschild-like spherically symmetric metric in polar-areal coordinates is given by
\begin{equation} \label{gen}
ds^{2}=-f(r)dt^{2}+\frac{1}{f(r)}dr^{2}+r^{2}d\theta^2+r^{2} \sin\theta d\phi^2, 
\end{equation}
where $f(r)$ is a general function of the radial coordinate $r$. In a spherical symmetric static background, the geodesic equation for a null particle can be found using Killing symmetries. Since the metric functions depend only on coordinates $r$ and $\theta$, there are two Killing vectors in this background: 
$K = \partial_t$ and $L = \partial_\phi$. These vectors describe the symmetry directions, and thus there are constants of motion associated with them. 
Those two constants of motion are the total energy, $E = f(r)\dot{t}$ and the angular momentum, $L = r^2 \dot{\phi}$. From these definitions, we can easily write
\begin{equation}  
r^2 \frac{d\phi}{dt}=f(r)b \quad\mathrm{with}\quad b\equiv\frac{L}{E}\, .
\end{equation}

Now, on the metric (\ref{gen}), we put the null geodesic condition $ds^2=0$, and for a null geodesic on the equatorial plane ($\theta=\pi/2$), we obtain
\begin{equation} \label{1st}
\frac{du}{d\phi}=\sqrt{\frac{1}{b^2}-u^2 f(u)}\, , 
\end{equation}
where $u\equiv\frac{1}{r}$. This equation shows us that the null geodesics depend on $b$, which is called the impact parameter for flat spacetimes. This is a equation for the null geodesic on the equatorial plane that contains only the first derivative of the function $u(\phi)$. If this equation can be solved for a specific $f(r)$ and $u(\phi)$ is determined, one can then evaluate the deflection angle by inverting the function $u(\phi)$.

If there exists a cosmological horizon in these coordinate systems, then we are interested in finding the coordinate angle difference for the motion of light from the cosmological horizon, $u_{h}$, at most to the closest approach distance, $u_{0}$. Otherwise, we would be calculating the coordinate angle difference for the motion between causally unconnected regions, which would be physically incorrect.

The closest point is defined by $\left. \frac{du}{d\phi} \right| _{u=u_0}=0$, and the cosmological horizon is defined by $f(r_h)=0$ with $u_h=1/r_h$. Thus, the coordinate angle difference is given by
\begin{equation}
 \bigtriangleup\phi\equiv\phi(r_{h})-\phi(r_{0})=\int_{u_{h}}^{u_{0}}\frac{du}{\sqrt{\frac{1}{b^2}-u^2 f(u)}}\, .
\end{equation}
From this, one finds the deflection angle as it travels from the source to the observer as
\begin{equation}
\bigtriangleup\alpha=2\bigtriangleup\phi-\pi \, ,
\end{equation}
which should be coordinate independent.

For Schwarzschild spacetime, $f(u) = 1-2mu$, and therefore the deflection angle is
\begin{equation}
\bigtriangleup\alpha=2\int_{u_{h}}^{u_{0}}\frac{du}{\sqrt{\frac{1}{b^2}-u^2+2m u^3}} - \pi 
\end{equation}
the exact solution in terms of an incomplete elliptic integral of the first kind of which was first given in Ref. \cite{Darwin1959} and much more recently in Ref. \cite{Batic2015}. In Ref. \cite{Batic2015}, the weak field limit (Eq. (33) of Ref. \cite{Batic2015}), which agrees with Ref. \cite{Keeton2005} and generalizes results of Refs. \cite{Virbhadra2000,Virbhadra1998}, and strong field limit (Eq. (40) of Ref.\cite{Batic2015}), which generalizes results of Refs. \cite{Darwin1959,Bozza2001,Bozza2002,Bisnovatyi2008}, of the deflection angle are also obtained by performing appropriate expansions of the first incomplete elliptic integral.

The square root in the null geodesic equation (\ref{1st}) is the reason for the complicated integral. We can get rid of the square root by writing the null geodesic equation in the form of a second-order ordinary differential equation (ODE) as
\be
\fr{d^2 u}{d\phi^2} = -\fr12 u^2 \fr{df}{du} -uf(u)\, .
\ee
The solution of any second-order ODE requires the specification of two boundary conditions. The boundary conditions that we choose define the point of closest approach of light to the lens: $u(0)=u_{0}$ and $u'(0)=0$. Note that these conditions define the closest point to the lens as the point with coordinates $\phi = 0$ and  $r = r_0 = b$. 

In the case in which the spacetime is nonflat, one has to apply perturbation methods to the flat spacetime solution, taking into account the cosmological horizon $u_{h}$. Then, the angle of deflection will be calculated for a special case of the intersection of the null geodesic and  the cosmological horizon.

%%%%%%%%%%%%%%%%%%%%%%%%%%%

\subsection{Weyl  gravity vacuum coordinates}\label{w}

Now, we find expression for the same deflection angle in a conformally related coordinate system. 
Using the radial coordinate transformation 
\begin{equation}
 \rho=\frac{r}{\sqrt{f(r)}}\left(\frac{\rho}{\rho_c}\right)^w
\end{equation}
we obtain a new coordinate system from (\ref{gen}) via a conformal transformation: 
\be
ds^2=\frac{\rho^2}{r^2}\left(-f(r)dt^{2}+\frac{1}{f(r)}dr^{2}+r^{2}d\theta^2+r^{2} \sin\theta d\phi^2\right)\, .
\ee
Therefore, we have a new conformal equivalent metric given by
\begin{equation}  
ds^{2}=-\left(\frac{\rho}{\rho_c}\right)^{2w} dt^{2}+\frac{1}{B(\rho)}d\rho^{2}+\rho^{2}d\theta^2+\rho^{2} \sin\theta d\phi^2\, ,
\end{equation}
with 
\begin{equation}
\frac{d\rho^2}{\rho^2 B(\rho)} = \frac{dr^2}{r^2 f(r)}\, .
\end{equation}
This is a kind of metric that can be written for the outer region of galaxies \cite{DKY2015} where the flat rotation curve phenomenon is observed. In that case, $\sqrt{w}$ is the rotating speed of a star moving on a circular orbit in the outer region of a galaxy. Thus, it is a very small number compared to speed of light, on the order of $10^{-3}$. Therefore, light sees the background described by this metric as if $w=0$. Hence, we take this parameter vanishing in the forthcoming calculations, and we use the new metric given by
\begin{equation}  \label{oth}
ds^{2}=-dt^{2}+\frac{1}{B(\rho)}d\rho^{2}+\rho^{2}d\theta^2+\rho^{2} \sin\theta d\phi^2\, .
\end{equation}  

This time from the Killing vector analysis, we find
\begin{equation}  
\rho^2\frac{d\phi}{dt}=b \quad\mathrm{with}\quad b\equiv\frac{L}{E}\, ,
\end{equation}
and after applying the null geodesic condition $ds^2=0$ on the metric (\ref{oth}), we obtain a first-order ODE for a null geodesic on the equatorial plane as 
\begin{equation} \label{1st-b}
\frac{dv}{d\phi}=\sqrt{(1-v^2 ) B(v)}\, , 
\end{equation}
where $v\equiv\frac{b}{\rho}$. Here, again, $b$ is the impact parameter for flat spacetimes. 

For the metric of interests in this paper, we will observe that there are no cosmological horizons in this new coordinate system. Therefore, the angle of deflection will be given by
\begin{equation} \label{da2}
\Delta\alpha=2\int^{1}_0 \frac{dv}{\sqrt{(1-v^2)B(v)}}-\pi\, ,
\end{equation}
where $1$ in the upper bound of the integral corresponds to the turning point $\rho = b$ and $0$ corresponds to the point at infinity.

We can get rid of the square root in the null geodesic equation by writing it in the form of a second-order ODE as
\be
\fr{d^2 v}{d\theta^2} = \fr12 (1-v^2 )\fr{dB}{dv}  -vB(v)\, .
\ee
The boundary conditions that we choose for the function $v(\theta)$ define the point of closest approach of light to the lens: $v(0)=v_{0}=1$ and $v'(0)=0$. Note that these conditions define the closest point to the lens as the point with coordinates $\theta = 0$ and  $\rho = \rho_0 = b$.

%%%%%%%%%%%%%%%%%%%%%%%%%%%

\section{Schwarzschild--de Sitter (Kottler) spacetime} \label{SdS}

\subsection{Polar-areal coordinates}

The metric in polar-areal coordinates is as given in (\ref{gen}) with 
\begin{equation}  
f(r)\equiv1-\frac{2m}{r}-\frac{\Lambda}{3}r^2\, .
\end{equation}
Then, the null geodesic equation as a second-order OPE becomes 
\begin{equation}
 \left(\frac{du}{d\phi}\right)^2=\frac{1}{b^2}+\frac{\Lambda}{3}-u^2+2mu^3, 
\end{equation}
where $u\equiv\frac{1}{r}$. Before solving this equation, we analyze turning points and horizons \cite{Frolov}.

Turning points are the points at which $ \frac{du}{d\phi}=0$. From this relation, one finds that
\be \label{tp}
az^3-z+1=0, 
\ee
where $z=r/2m$ and  $a=4m^2\left(1/b^2+\Lambda/3\right).$ There are just three cases to consider. In the following, we are going to write just the turning points and ignore negative and complex roots:
\begin{enumerate}
\item $0<a<4/27$ : there are two turning points given by
\beqa
z_{0} &=& \frac{2}{\sqrt{3a}}\cos\left(\frac{\pi-\Psi}{3}\right)\, ,\\
\mathrm{and}\quad z_{-} &=& \frac{2}{\sqrt{3a}}\cos\left(\frac{\pi+\Psi}{3}\right)\, ,
\eeqa
where $\cos^2\Psi=27a/4$.
\item $a=4/27$ : there is only one turning point given as
\be
z_{0}=z_{-}=z_{\gamma}=\frac{3}{2}\, .
\ee
Two turning points in the previous case approach together to coalesce at the photon sphere, located at $z_{\gamma}$.
\item $a>4/27$ : there are no turning points.
\end{enumerate}

Cosmological horizons are located at points where $f(r)=0$. This condition in the present case is equivalent to 
\be \label{ch}
yz^3-z+1=0\, ,
\ee
where $y=4m^2\Lambda/3$. Note that parameters $a$ and $y$ appear in Eqs. (\ref{tp}) and (\ref{ch}), respectively, in the same way. Thus, for the same range of values, a cosmological horizon or horizons exist: 

\begin{enumerate}
\item  For $0 < y < 4/27$, there are two horizons given by
\beqa
z_{c} &=& \frac{2}{\sqrt{3y}}\cos\left(\frac{\pi-\beta}{3}\right)\, ,\\
\mathrm{and}\quad z_{h} &=& \frac{2}{\sqrt{3y}}\cos\left(\frac{\pi+\beta}{3}\right)\, ,
\eeqa
where $\cos^2\beta=27y/4$, and we note that $z_{c} > z_{h}$. 
\item For $y=4/27$, there is only one horizon, but since it coincides with the photon sphere, it is useless for lensing calculations. 
\end{enumerate}

Thus, we calculate the deflection angle for the first case by using the variables and parameters $ v=r_{0}/r$, $\Lambda_{0}=\Lambda r_{0}^2$ and  $m_{0}=m/r_{0}^2$. In integral form, the deflection angle is
\begin{equation}
 \bigtriangleup\alpha=2\int_{v_{c}}^{1}\frac{dv}{\sqrt{1-v^2+2m_{0}v^3+\Lambda_{0}/3}}-\pi,
\end{equation}
where  $v_{c}=\frac{\sqrt{\Lambda_{0}}}{2} / \cos\left(\frac{\pi-\beta}{3}\right)$. Here, the integral can be evaluated after series expansion of the integrand to second order in $m_{0}$ and first order in $\Lambda_{0}$. The result (as previously found in Ref. \cite{Batic2015}) is
\be \label{Kpada}
\bigtriangleup\alpha = -2\sqrt{\frac{\Lambda_{0}}{3}}+ m_{0}\left(4-2\sqrt{\frac{\Lambda_{0}}{3}}-2\frac{\Lambda_{0}}{3}\right) 
+m_{0}^2\left(\frac{15}{4}\pi-4 \right) -m_{0}^2\left(3\sqrt{\frac{\Lambda_{0}}{3}}+2\frac{\Lambda_{0}}{3}\right) + \cdots 
\ee

%%%%%%%%%%%%%%%%%%%%%%%%%%%

\section{Mannheim-Kazanas spacetime} \label{MKS}

\subsection{Polar-areal coordinates}

The Weyl gravity solution in polar-areal coordinates was given long time ago by MK in Ref. \cite{MaKa} with
\begin{equation}  \label{MK}
ds^{2}=-f(r)dt^{2}+\frac{1}{f(r)}dr^{2}+r^{2}(d\theta^2+\sin^2\theta d\phi^2 ), 
\end{equation}
where
\be
f(r) =\sqrt{1-6m\gamma}-\frac{2m}{r}+\gamma r-k r^2 \, .
\ee    
Here, $m$ is the mass, $\gamma$ is the MK parameter, and $k$ is related to the cosmological constant by $k=\fr{\Lambda}3$. 

The null geodesic equation for this spacetime is given by
\begin{equation}
 \left(\frac{dr}{d\phi}\right)^2=r^4\left[\frac{1}{b^2}+\frac{\Lambda}{3}-\frac{\sqrt{1-6 m \gamma}}{r^2}+\frac{2m}{r^3}-\frac{\gamma}{r}\right]. 
\end{equation}
Before solving this equation, we analyze turning points and horizons as in the Kottler spacetime case.

Turning points are the points at which $ \frac{dr}{d\phi}=0$. From this relation, one finds that
\begin{equation}
\left(\frac{\Lambda}{3}+\frac{1}{b^2}\right)r^3-\gamma r^2-\sqrt{1-6 m \gamma}r+2m=0.
\end{equation}
There are again just three cases to consider. In the following, we write only the physically meaningful ones:
\begin{enumerate}
\item $0<m^2\left(\Lambda+\frac{3}{b^2}\right)\frac{\epsilon^2}{\eta^3}<1/9$ : there are three turning points given by 
\beqa
&r_{0}=2\sqrt{\frac{\eta}{\Lambda+\frac{3}{b^2}}}\cos{\frac{\pi-\Psi}{3}}+\frac{\gamma}{\Lambda+\frac{3}{b^2}}\, ,\\
&r_{+}=2\sqrt{\frac{\eta}{\Lambda+\frac{3}{b^2}}}\cos{\frac{\pi+\Psi}{3}}+\frac{\gamma}{\Lambda+\frac{3}{b^2}}\, ,\\
\mathrm{and}\quad &r_{-}=-2\sqrt{\frac{\eta}{\Lambda+\frac{3}{b^2}}}\cos{\frac{\Psi}{3}}+\frac{\gamma}{\Lambda+\frac{3}{b^2}}\, ,
\eeqa
where $\cos^2{\Psi}=9m^2\left(\Lambda+\frac{3}{b^2}\right)\frac{\epsilon^2}{\eta^3}$ with $\epsilon=1-\frac{\gamma^3}{3m\left(\Lambda+\frac{3}{b^2}\right)^2}-\frac{\gamma}{4m}\frac{\sqrt{1-6 m \gamma}}{\Lambda+\frac{3}{b^2}}$ and $\eta=\sqrt{1-6 m \gamma}+\frac{\gamma^2}{\Lambda+\frac{3}{b^2}}$.
\item When $\Psi=0$, the limit point of these turning points is the radius of the photon sphere ($r_{\gamma}$) given by
\begin{equation}
r_{0}=r_{+}=r_{\gamma}=\frac{1}{\gamma}(1-\sqrt{1-6m\gamma}).
\end{equation} 
\end{enumerate}
This result can be found easily using circular null geodesic conditions \cite{Cardoso2008}.

To find locations of cosmological horizons, we set $f(r)=0$, which is equivalent to
\begin{equation}
\frac{\Lambda}{3} r^3-\gamma r^2-\sqrt{1-6 m \gamma}r+2m=0.
\end{equation}
The physically meaningful solutions are found for $0<m^2\Lambda\frac{\varepsilon^2}{\xi^3}<1/9$. For this case, there are two turning points given by
\beqa
r_{c}=2\sqrt{\frac{\xi}{\Lambda}}\cos{\frac{\pi-\delta}{3}}+\frac{\gamma}{\Lambda}\, ,\\
\mathrm{and}\quad r_{h}=2\sqrt{\frac{\xi}{\Lambda}}\cos{\frac{\pi+\delta}{3}}+\frac{\gamma}{\Lambda}\, ,
\eeqa
where $\cos^2{\delta}=9m^2\Lambda\frac{\varepsilon^2}{\xi^3}$ with $\varepsilon=1-\frac{\gamma^3}{3m\Lambda^2}-\frac{\gamma}{4m}\frac{\sqrt{1-6 m \gamma}}{\Lambda}$ and $\xi=\sqrt{1-6 m \gamma}+\frac{\gamma^2}{\Lambda}$.

The geodesic equation can be written in terms of the turning point $r_{0}$ as
\beqa
\left(\frac{dv}{d\phi}\right)^2 &=& \sqrt{1-6 m_{0} \gamma_{0}}\ (1-v^2)+\gamma_{0}\ (1-v) +2m_{0}\ (v^3-1) \nonu \\
&=& 2m_{0}(1-v)(v_{+}-v)(v-v_{-}),
\eeqa
where $v=\frac{r_{0}}{r}$ and $v_{\pm}=\frac{1}{4m_{0}}\sqrt{1-6 m_{0} \gamma_{0}}-2m_{0}\pm \sqrt{1+2m_{0}\gamma_{0}+4m_{0}\sqrt{1-6 m_{0} \gamma_{0}}-12m_{0}^2}$.
The deflection angle is then given by
\beqa
 \bigtriangleup\alpha &=& 2\int_{v_{c}}^{1}\frac{dv}{\sqrt{2m_{0}(1-v)(v_{+}-v)(v-v_{-})}}-\pi \nonu \\
 &=& \frac{4}{\sqrt{2m_{0}(v_{+}-v_{-})}}F(p,q)-\pi \, ,
\eeqa
where $v_c=\frac{r_{0}}{r_c}$ and $F(p,q)$ is the elliptic integral of the first kind with $\sin{p}=\sqrt{\frac{(v_{+}-v_{-})(1-v_{c})}{(1-v_{-})(v_{+}-v_{c})}}$ and $q=\sqrt{\frac{1-v_{-}}{v_{+}-v_{-}}}$.
Using the expansion of $q$ in $\sqrt{m}$ as
\be 
q\approx \sqrt{2 \gamma +4} \sqrt{m}-\frac{3 \left(\sqrt{2} \gamma +\sqrt{2}\right) m^{3/2}}{\sqrt{\gamma +2}} + \cdots\, ,
\ee 
and the asymptotic expansion of $F(p,q)$ in $q$ given by
\be
F(p,q) \approx  p+q^2 \left(\frac{p}{4}-\frac{1}{8} \sin (2 p)\right) 
+\frac{3}{256} q^4 (12 p-8 \sin (2 p)+\sin (4 p)) + \cdots\, , 
\ee
we find the deflection angle as
\beqa \label{main}
\Delta\alpha &=& m_{0}\left(4-2\sqrt{\frac{\Lambda_{0}}{3}}-2\frac{\Lambda_{0}}{3}\right)
-2\sqrt{\frac{\Lambda_{0}}{3}}+\gamma_{0}\sqrt{\frac{\Lambda_{0}}{3}} 
+m_{0}^2\left(\frac{15\pi}{4}-4-3\sqrt{\frac{\Lambda_{0}}{3}}-2\frac{\Lambda_{0}}{3}\right) \nonu \\
&&+m_{0}\gamma_{0}\left(2+\frac{\Lambda_{0}}{3}\right) 
+m_{0}^2\gamma_{0}\left(\frac{15\pi}{4}-4-\frac{3}{2}\sqrt{\frac{\Lambda_{0}}{3}}\right) + \cdots
\eeqa

%%%%%%%%%%%%%%%%%%%%%%%%%%%

\subsection{Weyl  gravity vacuum coordinates}

Any solution of Weyl gravity should be conformally equivalent to the MK solution \cite{MaKa}. To show that the metric
\begin{equation}  
ds^{2}=-\left(\frac{\rho}{\rho_{c}}\right)^{2w} dt^{2}+\frac{1}{B(\rho)}d\rho^{2}+\rho^{2}(d\theta^2+\sin^2\theta d\phi^2 ) 
\end{equation}
is conformally equivalent to the MK metric (\ref{MK}), we write conformal equivalence condition as
\be\
\frac{\rho^2}{r^2}(-f(r)dt^{2} + \frac{1}{f(r)}dr^{2}+r^{2}d\Omega^2) =
-\left(\frac{\rho}{\rho_{c}}\right)^{2w} dt^{2} +\frac{1}{B(\rho)}d\rho^{2}+\rho^{2}d\Omega^2 \, , 
\ee
from which we obtain two equations:
\beqa 
\frac{\rho^2}{r^2}f(r) &=& \left(\frac{\rho}{\rho_{c}}\right)^{2w} \\
\frac{\rho^2}{r^2}\frac{1}{f(r)}\left( \frac{dr}{d\rho} \right)^2 &=& \frac{1}{B(\rho)}
\eeqa
From these equations, one finds
\begin{equation} \label{B}
B(\rho)=\frac{\rho^{2w-4}}{\rho_c^{2w}}\frac{1}{(du/d\rho)^2}
\end{equation}
and
\begin{equation} \label{v}
 2mu^3-\sqrt{1-6m\gamma}\ u^2-\gamma u +\fr{\Lambda}3+\frac{\rho^{2w-2}}{\rho_c^{2w}}=0,
\end{equation}
where $u\equiv\frac{1}{r}$. Solutions of the cubic algebraic equation (\ref{v}) are given by
\beqa
 u_1 &=& \frac{1}{6m}(h+h^{-1}+\sqrt{1-6m\gamma})\, ,\\
 u_2 &=& \frac{1}{6m}(e^{i4\pi/3}h+e^{i2\pi/3}h^{-1}+\sqrt{1-6m\gamma})\, ,\\
 u_3 &=& \frac{1}{6m}(e^{i2\pi/3}h+e^{i4\pi/3}h^{-1}+\sqrt{1-6m\gamma})\, ,
\eeqa
where
\be
 h(\rho) = \left[Q+\sqrt{Q^2-1}\right]^{1/3} ,\quad  Q(\rho) = [-\frac{\Delta_1}{\rho^{2(1-w)}}+\Delta_2] \, ,
 \ee
with
\be \label{delta}
\Delta_1\equiv\frac{54m^2}{\rho_c^{2w}} \quad \mathrm{and}\quad \Delta_2\equiv(1+3m\gamma)\sqrt{1-6m\gamma}-54m^2 \fr{\Lambda}3 \, .
\ee
 
Now, we can  find the metric component $B(\rho)$  from Eq. (\ref{B}). After some algebra, we find
\beqa
B_1(\rho) &=& \frac{3\rho^{2(1-w)}}{8(1-w)^2\Delta_1}(1+h^2+h^{-2})^2\, ,\\
B_2(\rho) &=& \frac{3\rho^{2(1-w)}}{8(1-w)^2\Delta_1}(1+e^{i2\pi/3}h^2+e^{i4\pi/3}h^{-2})^2\, ,  \\
B_3(\rho) &=& \frac{3\rho^{2(1-w)}}{8(1-w)^2\Delta_1}(1+e^{i4\pi/3}h^2+e^{i2\pi/3}h^{-2})^2\, . 
\eeqa
These are the solutions found in Ref. \cite{DKY2015} [Eq. (45) of that paper] by solving the field equations of Weyl gravity for a spherically symmetric metric. Thus, we reached these solutions from an alternative route, and this connects the present work with Ref. \cite{DKY2015}. Note that $ \Delta_{1,2}$ (\ref{delta}) correspond to the integration constants in the solutions of the field equations, $C_{1,2}$ in Ref. \cite{DKY2015}, respectively. We take the parameter $w$ vanishing in the forthcoming calculations as explained in Sec. \ref{w}.

We now need to evaluate the deflection angle integral given by
\begin{equation} \label{def}
\Delta\alpha=2\int^{1}_0 \frac{dv}{\sqrt{B (v)(1-v^2)}}-\pi.
\end{equation}
where $ v=\fr{\rho_0}\rho = \frac{b}{\rho}$ . To evaluate this complicated integral, we make the further redefinition that $\cos \zeta = Q(\rho)$, and after some algebra, we obtain
\begin{equation} \label{Br}
B_i(v)=\frac{1}{144m_{0}^2 v^2}[1+2\cos(\fr{2\zeta}3+(i-1)\fr{2\pi}3)]^2\, ,
\end{equation}
where $m_0\equiv\frac{m}{\rho_0}$.

To evaluate the integral (\ref{def}), we use $B_2(v)$. We first expand the integrand of (\ref{def}) perturbatively in terms of $m_0$, and then evaluating the integral, we obtain
\beqa \label{main1}
\Delta\alpha &=&4m_0-\pi+\frac{1}{(1+\nu^2)\nu}\left(\frac{45m_0^2\gamma_0^4}{16} - \frac{m_0\gamma_0^3}{2}\right)  
+ \frac{m_0^2\gamma_0^6}{16}\frac{1+3\nu^2}{\nu^3(1+\nu^2)^2}+2\sin^{-1}\frac{1}{\sqrt{1+\nu^2}} \nonu \\
&&+15m_0^2\left( \frac{\nu}{2}+\frac{1+\nu^2}{2}\sin^{-1}\frac{1}{\sqrt{1+\nu^2}} \right) + \cdots \, , 
\eeqa 
where $\gamma_0\equiv \gamma \rho_0$ and $\nu^2 = \fr{\Lambda_0}3+\fr{\gamma_0^2}{4}$ with $\Lambda_0\equiv \Lambda \rho_0^2$.

Expanding this expression first in $\gamma$ and then in $\Lambda$, we obtain
\begin{equation}
 \Delta\alpha=4m_0+\frac{15\pi}{4}m_0^2-2\sqrt{\fr{\Lambda_0}3}+\frac{15\pi}{4}m_0^2 \fr{\Lambda_0}3  + \cdots
\end{equation}
Note that this result is coordinate independent because $r_0=b$, where b is the impact parameter for flat spacetimes. After coordinate transformation,
\be \label{ct}
\rho_{0}=\frac{r_{0}}{\sqrt{f(r_{0})}}\, ,
\ee
the deflection angle in Mannheim-Kazanas coordinates is found to be
\beqa \label{main2}
\Delta\alpha &=& m_{0}\left(4-2\sqrt{\frac{\Lambda_{0}}{3}}-2\frac{\Lambda_{0}}{3}\right)
-2\sqrt{\frac{\Lambda_{0}}{3}}+\gamma_{0}\sqrt{\frac{\Lambda_{0}}{3}} 
+m_{0}^2\left(\frac{15\pi}{4}-4-3\sqrt{\frac{\Lambda_{0}}{3}}-2\frac{\Lambda_{0}}{3}\right) \nonu \\
&&+m_{0}\gamma_{0}\left(2+\frac{\Lambda_{0}}{3}\right) 
+m_{0}^2\gamma_{0}\left(\frac{15\pi}{4}-4-\frac{3}{2}\sqrt{\frac{\Lambda_{0}}{3}}\right) + \cdots \, ,
\eeqa
where  
\be
m_0\equiv\frac{m}{r_0} ,\ \gamma_0\equiv \gamma r_0 ,\, \mathrm{and}\, \Lambda_0\equiv \Lambda r_0^2.
\ee

Equation (\ref{main2}) for the deflection angle, which is equivalent to Eq. (\ref{main}) of the previous section, is our main result. We now look at two special cases: 1) the $\gamma = 0$ case to compare our result to a previous one \cite{Batic2015} obtained for the Kottler spacetime, which is equivalent to the MK spacetime for $\gamma = 0$, and 2) the $\Lambda = 0$ case to see the contribution of Weyl gravity to the bending of light in the Schwarzschild geometry.

%%%%%%%%%%%%%%%%%%%%%%%%%%%

\subsubsection{$\gamma = 0$ case} \label{gamma0}

To find the Kottler metric in Weyl  gravity vacuum coordinates, we take $\gamma = 0$ in (\ref{v}), and then the metric function $B(\rho)$ becomes
\begin{equation}
B(v)=\frac{1}{144m_{0}^2 v^2}[1+2\cos(\fr{2\zeta}3+\fr{2\pi}3)]^2\, ,
\end{equation}
where $\cos\zeta=1-54m_{0}^2(\Lambda_{0}/3+v^2)$ and $v=\rho_0/\rho$. In these coordinates, we also defined $m_{0}=m/\rho_{0}$ and $\Lambda_{0}=\rho_{0}^2\Lambda$. Since the distance of closest approach is $\rho_{0}=b$, the result will be coordinate independent. 

The deflection angle (\ref{da2}), after series expansion of the integrant to second order in $m_{0}$ and first order in $\Lambda_{0}$, is found to be
\begin{equation}
\bigtriangleup\alpha=-2\sqrt{\frac{\Lambda_{0}}{3}}+ 4m_{0}+\frac{15\pi}{4} m_{0}^2+\frac{15\pi}{4} m_{0}^2\frac{\Lambda_{0}}{3} + \cdots
\end{equation}
Using the coordinate transformation 
\be 
\rho_{0}=\frac{r_{0}}{\sqrt{f(r_{0})}}\, ,
\ee
we obtain the same result (\ref{Kpada}) as in the Kottler polar-areal coordinates. 

This result agrees with the Eq. (55) of Ref. \cite{Batic2015} for $\gamma_0 =0$. When comparing their result for the Kottler case with the prior literature, Batic \textit{et al.} \cite{Batic2015} first noted that it depends on the cosmological constant explicitly as first proclaimed in Ref. \cite{Rindler2007}. Next, they required that the weak field result for the gravitational deflection angle in Kottler spacetime should reproduce the weak field result for the Schwarzschild case in the limit of $\Lambda \rightarrow 0$. They noted that the latest result of Ishak and Rindler \cite{Rindler2010} does not agree with the weak field lensing limit of the Schwarzschild formula  (Eq. (33) of Ref.\cite{Batic2015}), which itself agrees with Ref. \cite{Keeton2005} and generalizes the results of Refs. \cite{Virbhadra2000,Virbhadra1998}. Agreement with the Schwarzschild case in the vanishing cosmological constant limit might be seen as an improvement over the previous literature, e.g., Refs. \cite{Ishak2008,Sereno2008,Schucker2009,Gibbons0808,Ishak2010,Nandi2011,Rindler2010,Bhadra2010}.

We would also like to comment on disagreements over the contribution of the cosmological constant $\Lambda$ to the deflection angle as argued in Refs. \cite{Khriplovich2008,Park2008,Arakida2012,Simpson2008}. In the Kottler spacetime, the physical impact parameter for light is given by $\fr1{B^2} = \fr1{b^2} + \fr{\Lambda}3$ \cite{Arakida2012,Lake2013,Lake2016}. Thus, when the orbit equation for light (\ref{1st}) is written in terms of $B$, it is observed that the null orbits are not influenced by $\Lambda$ \cite{Islam1983}. There is, of course, no disagreement on this fact in the literature. The main question is whether the finally calculated deflection angle or the lens equation depends on $\Lambda$ or not. As it is thoroughly analyzed in Ref. \cite{Lake2013} and concluded in Ref. \cite{Lake2016},  $\Lambda$ appears in the final lens equation when it is expressed in terms of measurable quantities (see discussion after Eq. (38) of Ref. \cite{Lake2016}). The negative results of Refs. \cite{Khriplovich2008,Park2008} are due to mixed usage of coordinate and measurable quantities in the final lens equation \cite{Lake2013}, whereas the main problem of Ref. \cite{Arakida2012} is  not properly taking into account the finite radius of the cosmological horizon of the Kottler spacetime \cite{Ishihara2016}. Simpson \textit{et al.}'s conclusion in \cite{Simpson2008} is similarly influenced by the incorrect characterization of the vacuole radius as pointed out in Refs. \cite{Ishak2010,Rindler2010,He2017}. 

%%%%%%%%%%%%%%%%%%%%%%%%%%%

\subsubsection{$\Lambda = 0$ case}

In the case in which $\Lambda =0$, our main result (\ref{main2}) for the deflection angle becomes 
\be \label{Coco0}
\Delta\alpha = 4m_{0} + 2m_{0}\gamma_{0} +m_{0}^2 (1+ \gamma_{0}) \left(\frac{15\pi}{4}-4\right)\, ,
\ee
up to $m_0^2$ and $\gamma_0$ order. 

This result shows that the MK parameter contributes positively to the deflection angle. Comparing our result to the existing ones in the literature we note that we agree, up to this order, with the result of Ref. \cite{Lim-Wang2017} [Eq. (35) of that paper]. 

If we also compare the first-order correction in $\gamma$ to the general relativistic result, we note that it has a piece independent of the impact parameter and a piece inversely proportional to the impact parameter, which is similar to the Schwarzschild contribution. Hence, these contributions increase the lensing effect of a galaxy cluster as expected from a theory alternative to dark matter phenomenology.

%%%%%%%%%%%%%%%%%%%%%%%%%%%

\subsubsection{Comparison with other works}

There are three kinds of first-order corrections in $\gamma$ to the general relativistic result \cite{Cutajar1403,Lim-Wang2017} in the literature: 

1) $\gamma_0 =\gamma \rho_0$ with negative sign in Refs. \cite{Walker1994,Edery9708,Nandi2010,Nandi2011} and with positive sign in Ref. \cite{Potapov2016}. This is clearly a wrong result because for a gravitational lens the deflection angle diminishes with the impact parameter contrary to what this result suggests \cite{Sultana2010}. It is possible to regain this result in our approach via a ``wrong'' order of expansion: expanding (\ref{main1}) first in $\Lambda$ and then in $\gamma$, we obtain
\be
\Delta\alpha = 4m_{0} -\gamma_{0} + m_{0}\gamma_{0} +m_{0}^2 \left(\frac{15\pi}{4}-4\right) 
+ m_{0}^2 \gamma_{0} \left(\frac{15\pi}{4}-\frac{11}2\right)\, ,
\ee
after performing coordinate transformation (\ref{ct}) and setting $\Lambda =0$.
 
We note that many works \cite{Nandi2010,Nandi2011,Sultana2010,Sultana2013,Cattani2013,Cutajar1403,Potapov2016,Lim-Wang2017} use the Rindler-Ishak bending formula (Eq. (16) in Ref. \cite{Rindler2007}) to determine the deflection of light. This formula contains $\Lambda$, which is contributed only by the metric function $f(r)$. The equation for the null geodesic is independent of $\Lambda$ and thus does not contribute the bending angle formula any further terms that include $\Lambda$. Since $\Lambda$ appears only in linear order in the bending formula of Rindler and Ishak \cite{Rindler2007}, the formula does not allow the choice of the ``correct'' order of expansion, i.e., first in $\gamma$ and then in $\Lambda$. We believe this is the reason for the unacceptable negative contribution of $\gamma_0$ to the bending angle in Refs. \cite{Nandi2010,Nandi2011}. In Refs. \cite{Walker1994,Edery9708}, $\Lambda$ was dropped out of the calculation too early, and thus there was again no choice of expansion in $\Lambda$. The reason that Refs. \cite{Sultana2010,Sultana2013,Cattani2013,Cutajar1403,Potapov2016} obtain a different contribution of $\gamma$ compared to Refs. \cite{Nandi2010,Nandi2011} is that the authors either keep (note the form of the metric function in Refs. \cite{Sultana2010,Sultana2013,Cattani2013,Cutajar1403}) or ignore (note the difference between Eqs. (47-48) and (52-53) in Ref. \cite{Potapov2016}) $\gamma$-dependent terms during the computation.

2) $m_{0}^2 \gamma_{0}$ of Refs. \cite{Sultana2013,Cattani2013,Cutajar1403} exists in our formula (\ref{Coco0}), but it is in second order in mass. The reason for this correction is explained in the above item.

3) $m_{0} \gamma_{0}$ of Ref. \cite{Lim-Wang2017} also exists in our formula (\ref{Coco0}). It is interesting to note that the definition of the deflection angle in Ref. \cite{Lim-Wang2017} is different than our definition (\ref{1st}). The effect of different definitions is not observed in the contribution of the MK parameter $\gamma$ but in the full deflection angle formula [compare Eq. (37) of Ref. \cite{Lim-Wang2017} with our main result (\ref{main2})]. Even though the authors also use the Rindler-Ishak bending angle formula, the final result of Ref. \cite{Lim-Wang2017} for the bending angle is drastically different than the results of Refs. \cite{Nandi2010,Nandi2011,Pireaux0408,Sultana2010,Sultana2013,Cattani2013,Cutajar1403,Potapov2016}. This is due to the fact that the location of the cosmological horizon, the radius of which is strongly dependent on the cosmological constant $\Lambda$, plays a very important role in their approach. This brings not just linear but complicated dependence on $\Lambda$, which, due to correlation with $\gamma$, causes a very different expression for the bending angle to be found when a series expansion in $\gamma$ is performed.

%%%%%%%%%%%%%%%%%%%%%%%%%%%

\section{Conclusions \label{discussion}}

We utilized our solution of the Weyl gravity \cite{DKY2015} to calculate the deflection angle of light from a distant source by a galaxy cluster. We first observed that our Weyl gravity solution is conformally equivalent to the MK solution, as any solution to Weyl gravity field equations should be. So, we have a different conformal factor compared to the MK solution, which makes difference only in the case of massive particle trajectories. Light trajectories do not distinguish conformally equivalent metrics; thus, our result is relevant for the discussion on the sign and the value of the MK parameter $\gamma$. Our calculation of the deflection angle in the Kottler spacetime included also the contribution of the cosmological constant $\Lambda$. This contribution, however, came out rather differently than the previous works in the literature. The reason of this difference comes from realization that how $\Lambda$ and also $\gamma$ contribute depends strongly on how and at what point in the calculation the perturbative expansions in various quantities are performed. These quantities are the mass $m$ of the gravitational lens, the cosmological constant $\Lambda$, and the MK parameter $\gamma$. In which order the perturbative expansions are made is very important. We found out that expansions first in $m$, then in $\gamma$, and finally in $\Lambda$ is the mathematically correct process because otherwise one gets higher-order terms larger than the lower-order terms  in the expansions. This is against the whole idea of perturbation expansion. We found that our result without the MK parameter for the deflection angle agrees with the analysis done in Ref. \cite{Batic2015}. Our result with the MK parameter also differs from the ones in the literature, partially agreeing only with the result presented in Ref. \cite{Lim-Wang2017}. We still have to check these formulas by using different methods such as Refs. \cite{Gibbons2008, Arakida2017} and then try to analyze observational data to see if our formula agrees at all with the observations without invoking dark matter. These are two ideas for future research and are beyond the scope of this paper.

%%%%%%%%%%%%%%%%%%%%%%%%%%%

\acknowledgments

C.D. thanks Toshitaka Kajino and the COSNAP group at NAOJ for discussions. O.K. was supported by TUBITAK-BIDEB 2211-A National scholarship program for PhD students. C.D. is supported in part by ICTP-SEENET-MTP project NT-03 Cosmology-Classical and Quantum Challenges.

%%%%%%%%%%%%%%%%%%%%%%%%%%%

\end{document}